# Ion Beam Synthesis of Embedded SiC

Y. S. Katharria[1*], V. Baranwal[2], D. C. Agarwal[3], M. Kumar[2], R. Krishna[2],

P. Kumar[1], F. Singh[1], D. Kanjilal[1]

[1] Nuclear Science Centre, Aruna Asaf Ali Marg, P. O. Box-10502, New Delhi-67

[2] Department of Physics, University of Allahabad, Allahabad-211002

[3] Department of Physics, R. B. S. College, Agra-282 002

**Abstract**

The synthesis of embedded silicon carbide (SiC) was carried out in n-type silicon (Si) samples having (100) and (111) orientations using high dose implantation of 150 keV carbon ($C^+$) ions at room temperature. The dose is varied from $4\times10^{17}$ ions-cm$^{-2}$ to $8\times10^{17}$ ions-cm$^{-2}$ in different samples of both the orientations. Post implantation annealing at 1000°C was done in order to anneal out the implantation induced defects and to get silicon carbide precipitates embedded in the silicon matrix. Detailed Fourier transform infrared spectroscopy analysis and x-ray diffraction studies confirm the formation of cubic phase (β-SiC). The grain size of the SiC precipitates is estimated to be a few nano-meters. The SiC precipitates have been found to exhibit a better crystalline order in Si (100) samples than in Si (111) samples. The x-ray diffraction results also indicate the amorphization of bombarded region of the Si samples. After annealing the amorphized region got recrystalized into a polycrystalline phase of Si.

## Introduction

Silicon carbide (SiC) has been of current interest due to its promising properties such as wide band gap, high breakdown electric field strength, high saturation drift velocity and high chemical and radiation hardness. These properties make SiC a suitable material for applications in semiconductor devices, which can be operated in extreme conditions of temperature, power and radiation [1-4]. Among the various techniques to grow SiC, ion beam synthesis (IBS) by $C^+$ implantation into Si substrates has been an attractive technique due to the precise control over the depth and dose of implanted ions, as well as the high purity of buried layers that is obtained. The process of IBS of SiC has long been explored for various aspects of its growth since the first such work proposed by *J. A. Boarder et al.* [5].

The aim of the aforementioned work is to present a systematic study of SiC nano-crystallites formation in the Si (100) and Si (111) matrices by $C^+$ ion implantation. X-ray diffraction (XRD) and Fourier transform infrared spectroscopy (FTIR) have been used to analyze the synthesis process. The effects of implanted $C^+$ dose and orientation of Si substrates on IBS of SiC have been explored.

## Experimental Details

N-type Si (100) and Si (111) samples were implanted with 150 keV $C^+$ ions to doses of $4\times10^{17}$, $5\times10^{17}$, $6\times10^{17}$ and $8\times10^{17}$ ions-cm$^{-2}$ using Low Energy Ion Beam Facility (LEIBF) at Nuclear Science Centre, New Delhi. Before implantation samples were cut into approximately 10mm×10mm size. The samples were chemically cleaned using standard procedure and passivated using 2% HF. During implantation, residual gas pressure in the experimental chamber was ~$5\times10^{-7}$ Torr and the beam line pressure was

~$3\times10^{-7}$ Torr. The $C^+$ ions were extracted from Electron Cyclotron Resonance (ECR) ion source and accelerated to energy of 150 keV. The beam current was ~5 µA during the experiment. After implantation, one set of samples was annealed at 1000°C for 30 minutes in the argon ambient in a tubular furnace. The phase formation of SiC in Si substrate was characterized by XRD performed using a Brooker AXS diffractometer in thin film mode keeping a step size of 0.020° and a step time of 10 seconds. The samples were further characterized by FTIR spectroscopy using Nexus 670 FTIR spectroscope to get the dose and orientation dependence on SiC growth.

**Results and Discussion**

The XRD pattern obtained for Si (100) and Si (111) samples implanted to a dose of $6\times10^{17}$ ions-cm$^{-2}$ and annealed at 1000°C are shown in figure 1 and figure 2, respectively. The x-ray peaks at $2\theta$ = 35.64° and 60.18° in these patterns correspond to reflections from (111) and (220) planes of SiC crystallites in cubic phase (β-SiC). The absence of reflections from other planes indicates epitaxial growth of SiC crystallites [6]. The x-ray peak at $2\theta$ = 35.64°, can be fitted using two peaks Gaussian fit (figure 1). One of the peaks is centered at 34.33° and the second one at 35.64°. As mentioned earlier the peak centered at 35.64° is due to SiC (111) reflection. The second peak can be due to the presence of a small number of SiC crystallites in hexagonal phase (e.g. (101) reflections of 6H-SiC and 4H-SiC crystallites). The strain in the SiC crystallites resulting from approximately 20% mismatch between lattice constants of SiC and Si unit cells can also contribute to this peak.

The size of β-SiC crystallites is calculated using Scherrer's formula:

$$t = \frac{0.9\lambda}{\Gamma \cos\theta} \quad \quad \quad (1)$$

Where, t (Å) is the size of the crystallites, λ (Å), the wavelength of the incident x-rays (1.54 Å), Γ (radians), the FWHM of the XRD peak and 2θ, the Bragg angle.

The size is found to be approximately 9 nm for Si (100) samples and 8 nm for Si(111) samples. The ratio of integrated intensities of peaks from SiC (111) reflection for Si (100) and Si (111) samples is found to be ~1.2 which indicates the dominance of growth of SiC crystallites in Si (100) samples than in Si (111) samples. The XRD pattern of as-implanted samples indicates the amorphization of implanted regions of Si samples, which gets recrystalized after annealing at 1000°C for 30 minutes.

FTIR spectra of the Si (100) samples implanted to a dose level of $6 \times 10^{17}$ ions-cm$^{-2}$ and annealed at 1000°C are shown in figure 3. The spectra of as-implanted samples are characterized by a broad absorption dip centered around 735 cm$^{-1}$. This absorption dip is assigned to SiC precipitates in amorphous phase [5]. After annealing, this absorption dip becomes narrower and shifts to higher energy side (~794 cm$^{-1}$). This absorption dip, observed in the FTIR spectra of the annealed samples is attributed to the transverse optical (TO) phonon absorption of crystalline SiC (c-SiC) in cubic and hexagonal phase [7,8]. An additional peak centered around 620 cm$^{-1}$ is due to asymmetric vibrations of Si lattice. Similar absorption bands are also observed for Si (111) samples prepared by implantation with the same doses as those for Si (100) samples.

It is obvious from figure 3 that the absorption dip assigned to c-SiC becomes more pronounced as the dose is increased from $4 \times 10^{17}$ ions-cm$^{-2}$ to $8 \times 10^{17}$ ions-cm$^{-2}$. These dips observed in samples implanted to different doses were fitted with Lorentzian peak fit and the area under peak is plotted with dose in figure 4. The area that is a measure of amount of SiC precipitates, increases monotonically with dose up to a dose of $6 \times 10^{17}$ ions-cm$^{-2}$ and almost saturates for a higher dose of $8 \times 10^{17}$ ions-cm$^{-2}$. An increase in the area under peak is expected due two reasons: firstly the higher amount of SiC can be precipitated out with higher number of $C^+$ being available and secondly ion beam induced crystallization (IBIC). The saturation of area observed for both Si (100) and Si (111) samples can be explained in terms of two competing processes of ion beam induced amorphization (IBIA), which is more pronounced for higher doses and ion beam induced crystallization (IBIC).

As reported earlier [9, 10] higher the FWHM of the FTIR absorption dip, smaller is the SiC grain size and vice versa. Taking this fact in to account, the area covered by the SiC absorption dips is divided by the FWHM of the corresponding dips. This quantity (area divided by FWHM), which is now independent of FWHM may be taken as a measure of the amount of the SiC precipitates with long-range order. The variation of this quantity (in relative units) with dose is shown in figure 5. As it is obvious from this figure, area/FWHM increases with increasing dose up to a dose of $6\times10^{17}$ ions-cm$^{-2}$ and then decreases slightly for the dose of $8\times10^{17}$ ions-cm$^{-2}$. This indicates an increasing amount of comparatively long-range SiC precipitates and can be due to IBIC. The decrease in area/FWHM can be attributed to IBIA which is dominant over IBIC for the doses above a critical level. The area/FWHM was found to be more for Si (100) samples then for Si (111) samples indicating a longer-range order of SiC crystallites in Si (100) substrates than in Si (111) substrates. This is also obvious from XRD size calculations and the ratio of integrated intensities of SiC (111) peaks for the two substrates. A further understanding of crystallinity of SiC precipitates in the substrates with (100) and (111) orientations can be gained from the damping constant ($\gamma$) [11] that is defined according to the relation:

$$\frac{1}{T(\omega)} - 1 = f(\omega/\omega_{TO}) \quad \ldots\ldots\ldots\ldots (2)$$

Where, $T(\omega)$ is the experimental transmittance. The FWHM of the function $f(\omega/\omega_{TO})$ is defined as the damping constant ($\gamma$). This factor is a measure of damping of optical vibrational modes and hence a measure of crystallinity of the environment [11]. The variation of damping constant with $C^+$ dose is plotted in figure 6 for Si samples of both the orientations. It is seen from figure 6 that the value of damping constant is higher for implanted Si (111) samples than for Si (100) samples, indicating a better crystallinity of SiC precipitates in Si (100) samples.

**Conclusions**

Room temperature implantation of 150 keV $C^+$ ions at doses varying from $4\times10^{17}$ ions-cm$^{-2}$ to $8\times10^{17}$ ions-cm$^{-2}$ followed by annealing at 1000°C has led to the formation of β-SiC nano-phase in Si (100) and Si (111) substrates. The Si (100) substrate is found to show a longer-range order of SiC crystallites and thus, higher growth of SiC crystallites as compared to that in Si (111) substrates. The crystallinity of SiC precipitates is better in Si (100) samples than in Si (111) samples.

**Acknowledgment**

One of the authors (YSK) is grateful to Council of Scientific and Industrial Research (CSIR) for providing fellowship for this work under CSIR award no. F.No.2-56/2002(I) EU.II. The authors are also thankful to Mr. G. Rodrigue for providing good quality stable beam.

**Figure Captions**:

**Figure 1**: XRD pattern for Si (100) samples implanted to a dose of $6\times10^{17}$ ions-cm$^{-2}$ (a) as-implanted (b) implanted and annealed. Insert shows the two peak Gaussian fit of SiC (111) peak.

**Figure 2**: XRD pattern for Si (111) samples implanted to a dose of $6\times10^{17}$ ions-cm$^{-2}$ (a) as-implanted (b) implanted and annealed. Insert is shows the two peak Gaussian fit of SiC (111) peak.

**Figure 3:** FTIR spectra of Si (100) samples implanted to varying doses from $4\times10^{17}$ ions-cm$^{-2}$ to $8\times10^{17}$ ions-cm$^{-2}$ (i) as-implanted, (ii) implanted and annealed.

**Figure 4:** Dependence of amount of SiC precipitate characterized by area under peak in FTIR spectra on implanted C$^{+}$ dose for Si (100) and Si (111) samples.

**Figure 5:** Dose Dependence of periodic order of SiC precipitates characterized by area of FTIR absorption band divided by FWHM of that band.

**Figure 6:** The variation of damping constant with implanted carbon dose in Si(100) and Si (111) samples.

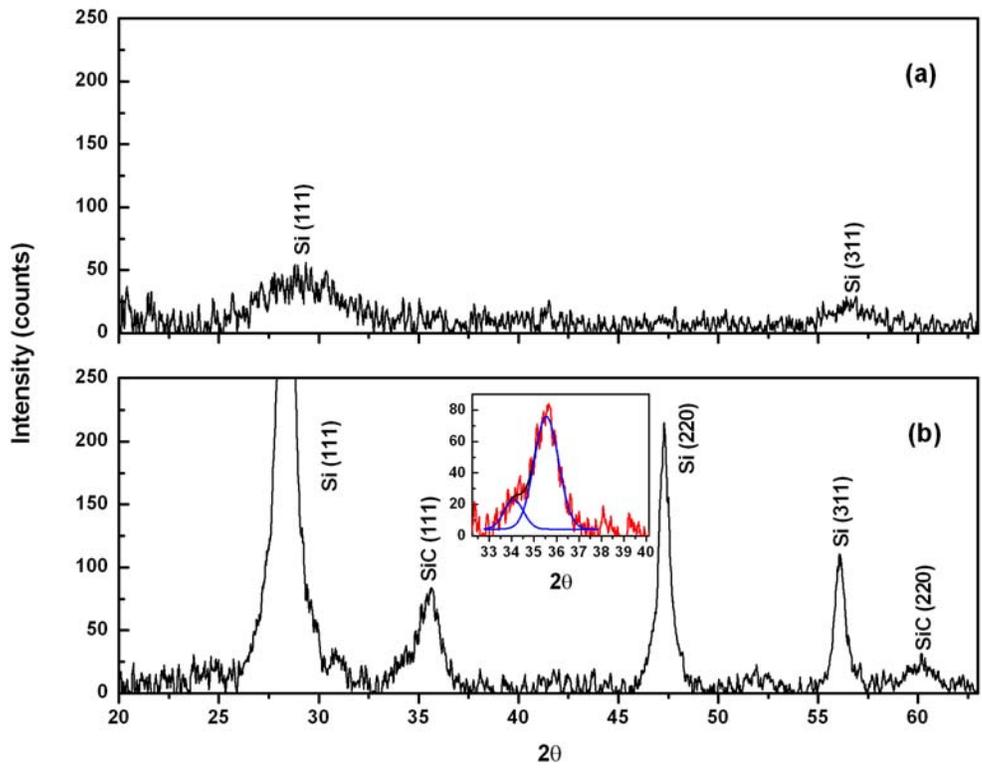

**Figure 1**

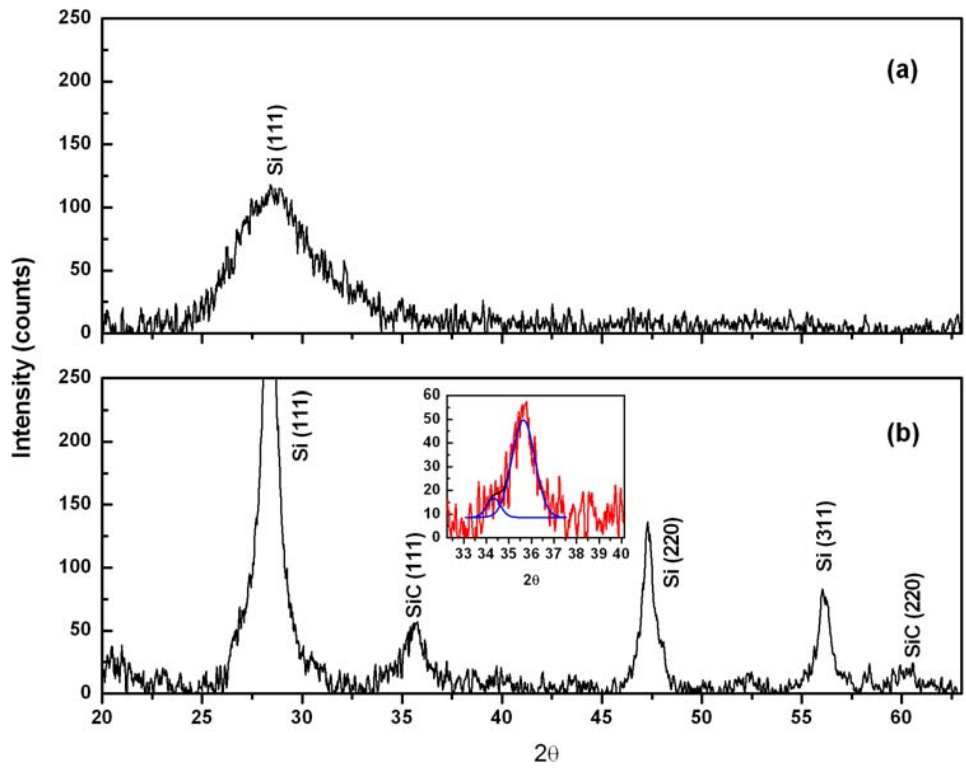

**Figure 2**

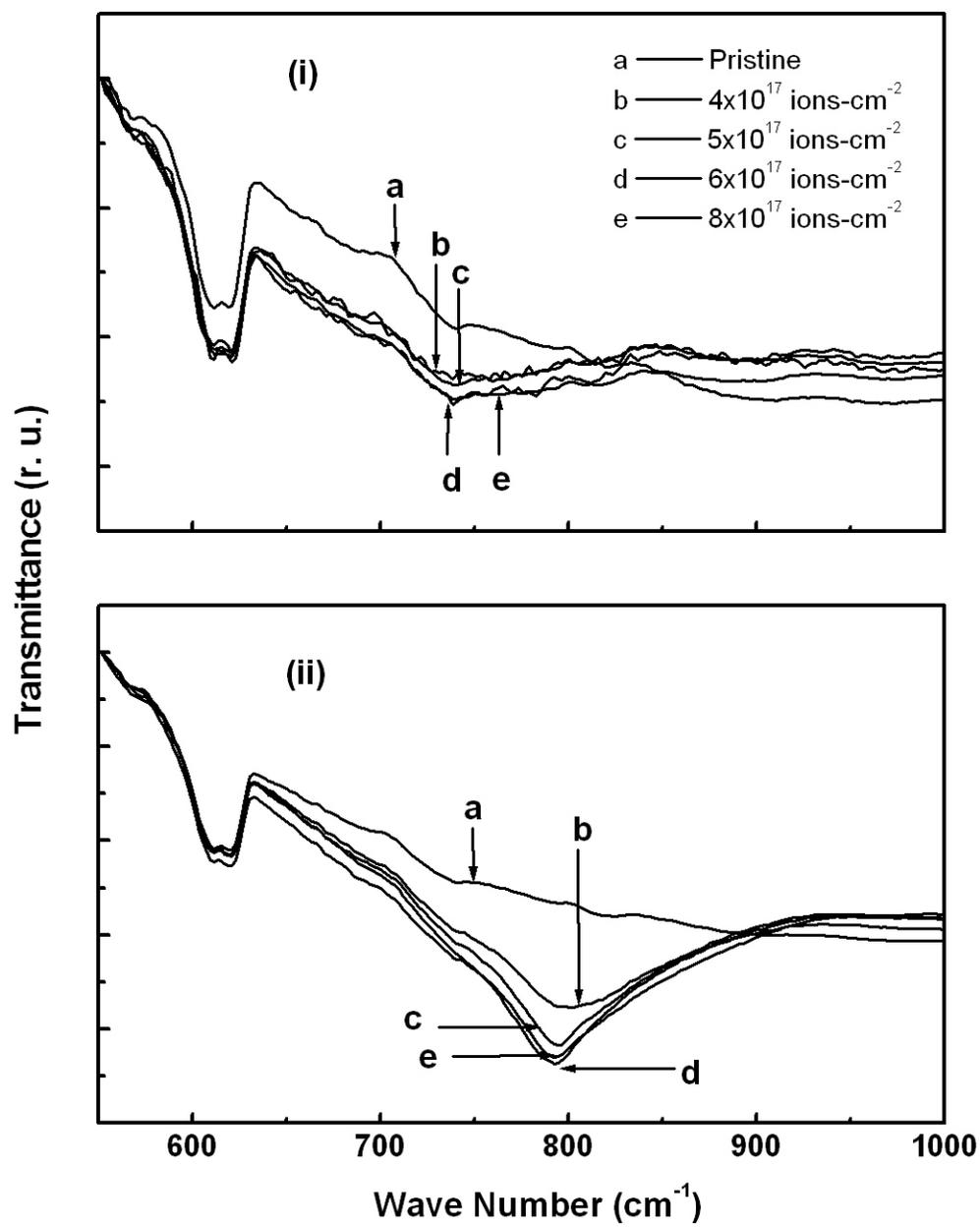

**Figure 3**

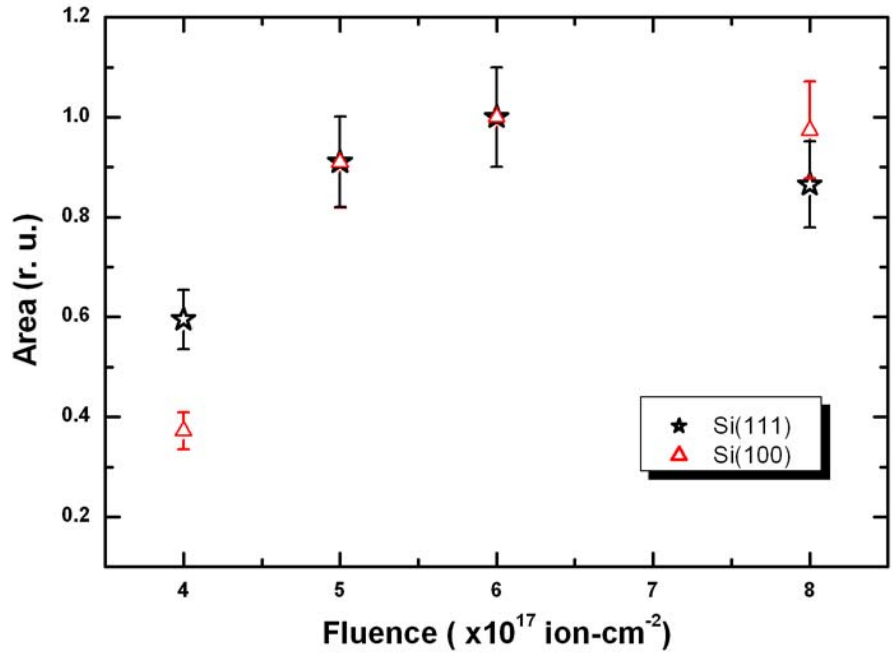

**Figure 4**

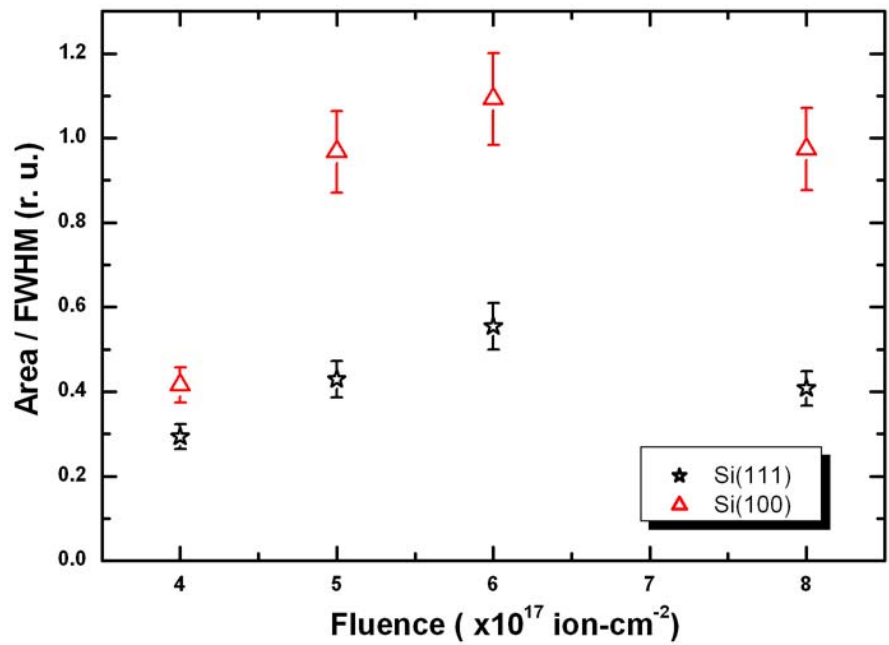

**Figure 5**

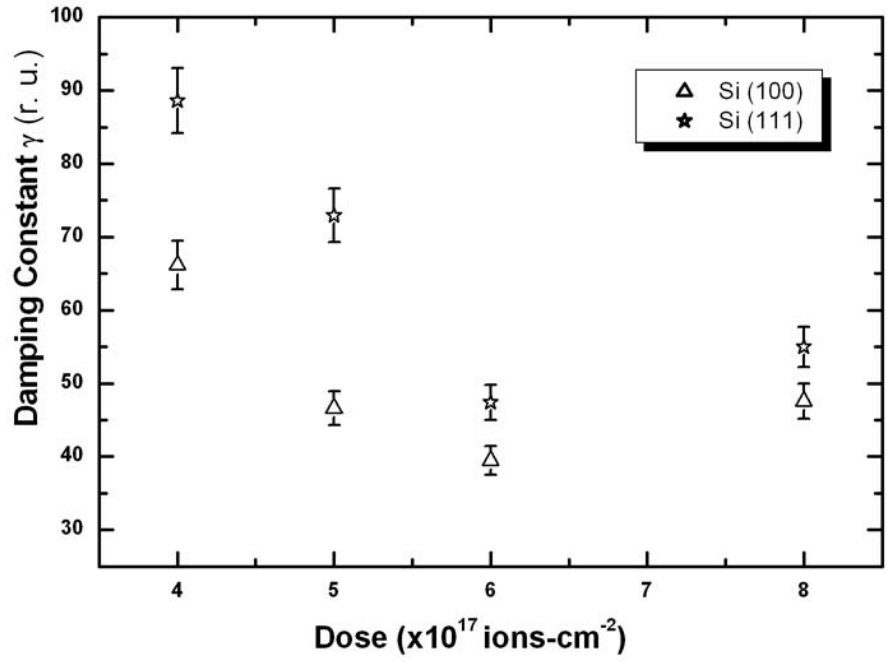

**Figure 6**